# The Power of Smartphones

**Feng Xia [1], Ching-Hsien Hsu [2], Xiaojing Liu [1], Haifeng Liu [1], Fangwei Ding [1], Wei Zhang [1]**

[1]School of Software, Dalian University of Technology, Dalian 116620, China

[2]Department of Computer Science and Information Engineering, Chung Hua University, Taiwan

**Abstract:** Smartphones have been shipped with multiple wireless network interfaces in order to meet their diverse communication and networking demands. However, as smartphones increasingly rely on wireless network connections to realize more functions, the demand of energy has been significantly increased, which has become the limit for people to explore smartphones' real power. In this paper, we first review typical smartphone computing systems, energy consumption of smartphone, and state-of-the-art techniques of energy saving for smartphones. Then we propose a location-assisted Wi-Fi discovery scheme, which discovers the nearest Wi-Fi network access points (APs) by using the user's location information. This allows the user to switch to the Wi-Fi interface in an intelligent manner when he/she arrives at the nearest Wi-Fi network AP. Thus we can meet the user's bandwidth needs and provide the best connectivity. Additionally, it avoids the long periods in idle state and greatly reduces the number of unnecessary Wi-Fi scans on the mobile device. Our experiments and simulations demonstrate that our scheme effectively saves energy for smartphones integrated with Wi-Fi and cellular interfaces.

**Keywords:** smartphone; power consumption; access point discovery; energy saving

## 1. Introduction

Smartphones are becoming increasingly intertwined with people's daily lives. They are now equipped with more powerful processors, more memory, multiple network interfaces, and more powerful operating systems such as Windows Phone, Google Android, Apple iOS, and so forth. Furthermore, the cellular networks of smartphones have grown from GSM networks to the current 3G networks, which have greatly increased the bandwidth for smartphones.

Today's smartphones are capable to support a large spectrum of applications from stock tickers to city-wide social games [1]. There are many new innovative applications about using smartphones as mobile sensors to recognize user state [2], sense urban space [3, 4], sense location [5], monitor environment conditions [6], monitor traffic [7], sense people's health states [8], and so on.

However, many applications such as mobile gaming and real-time location-based tracking applications [9] consume significant battery energy because of the requirement of continuous internet connection [10]. However, the increased size of battery hasn't kept up with the increased demand for energy. Advances in battery technology are growing slowly and have not kept pace with rapidly growing energy demands [11]. Thus the power consumption has become an important problem of the energy management of smartphones. Furthermore, energy saving for smartphones has become an urgent matter, which is beneficial to extending smartphones' standby time and battery's lifetime as well as reducing the number of times of battery recharging.

Many wireless network technologies, such as Bluetooth, Wi-Fi, and GPRS, have become popular in the market. Wi-Fi and GPRS have different characteristics in terms of power, range and bandwidth [12, 13]. GPRS has low transfer power efficiency[1], wide range, and low bandwidth. On the contrary, Wi-Fi has high transfer power efficiency, narrow range, and high bandwidth. However, Wi-Fi has high power consumption in idle state and brings a high overhead when scanning for new networks.

Mobile devices, such as smartphones, are equipped with multiple network interfaces that have complementary characteristics, as mentioned before. This brings a new challenge to the choice the most suitable network interface to connect to the network. Therefore we should leverage the complementary characteristics of these network interfaces to meet the user's bandwidth needs, provide the best connectivity and minimize the power consumption at the same time. Many researchers (e.g. [14, 15]) have worked on saving energy by reducing the energy costs of network interfaces.

In this paper, we address this challenge by presenting a simple scheme that discovers the nearest Wi-Fi network access point (AP) by using the user's location information. Thus users can switch to the Wi-Fi network when they approach it, which greatly reduces the number of unnecessary Wi-Fi scans on the mobile devices. As a result, much energy can be saved.

Currently, many location sensing technologies, such as GPS, GSM and ZigBee, have been developed for mobile devices. GPS, GSM and ZigBee have widely different characteristics in terms of power and accuracy (see Table 1) [15, 16]. GPS has high positioning accuracy, but it is power-demanding and can be used only in outdoors. GSM is power efficient, but it is highly inaccurate. ZigBee is energy efficient and has high positioning accuracy, but mobile devices with ZigBee interfaces is still very scarce and expensive. In addition, ZigBee is mostly used in indoors. Assisted GPS (A-GPS for short) is the same as GPS in terms of accuracy and cost less power, and only used in outdoors. Considering energy-accuracy trade-off and restrictions of different location sensors, we adopt A-GPS in outdoors and ZigBee in indoors to get user's location information.

In terms of contributions, we make efforts in the following aspects. First, we review typical smartphone's sensing applications and systems. Second, we analyze and test the energy consumption of smartphones. Third, main approaches of energy saving for smartphones are summarized. Finally, we propose a location assisted Wi-Fi discovery scheme that greatly reduces the number of unnecessary Wi-Fi scans on mobile devices. Our simulations and experiments demonstrate that it is effective.

**Table 1.** Characteristics of various location sensing technologies.

| Sensor | Average power consumption (mW) | Approximate accuracy (m) |
|--------|-------------------------------|--------------------------|
| GPS | 400 | 10 |
| GSM | 60 | 400 |
| ZigBee | 1 | 5 |

The rest of the paper is structured as follows. Section 2 describes the "power" of smartphones: smartphones as sensors in typical sensing applications and systems. The "power"/energy consumption of smartphones and their diverse wireless network interfaces is analyzed and tested in Section 3. We summarize the main energy-saving techniques for smartphones in Section 4. Section 5 shows the key components and operation workflow of our proposed system. In Section 6, we evaluate the power consumption of our five outdoor schemes, and the evaluation of five indoor schemes is done in Section 7. We conclude the paper in Section 8.

---

[1] Power efficiency: with the same power to use, the more data that can be transferred under a scheme, the higher power efficiency it has. Here we mean that, with the same energy consumption, GPRS transfer less data than Wi-Fi, so the Wi-Fi has higher power efficiency than GPRS.

## 2. Smartphones as Sensors

Mobile phones not only serve as the key computing and communication mobile device of choice, but also come with a rich set of embedded sensors, such as accelerometer, compass, GPS, gyroscope, microphone, and camera. Recently, sensors have become much more prevalent in mobile devices like smartphones. Smartphones serve our life across a wide variety of domains, such as health monitoring, social networking, environmental monitoring, traffic monitoring, and human behavioral monitoring, and give rise to a new area of research called phone sensing. Many mobile phone sensing systems have been developed in the literature. In this section, we survey these systems according to their application domains, as outlined in Table 2.

**Table 2.** Mobile sensing systems.

| Area | System | Type(s) of Sensors | Application(s) |
|------|--------|--------------------|----------------|
| Health Monitoring | SPA [17] | Biomedical sensor, GPS | Healthcare suggestions |
| | UbiFit Garden [18] | 3D accelerometer | UbiFit Garden's interactive application |
| | BALANCE [19] | Accelerometer, GPS | BALANCE |
| | CONSORTS-S [20] | Wireless sensor, MESI RF-ECG | Healthcare service |
| Environmental Monitoring | PEIR [21] | GPS | Carbon impact, Sensitive site impact, Smog exposure, etc |
| | NoiseTube [22] | Microphone, GPS | Mobile sensing |
| | MobGeoSen [23] | Camera, GPS, Microphone | Sound level monitoring |
| | Laemometer [24] | Microphone, GPS | Noise map creation and visualization |
| Traffic Monitoring | V-Track [25] | GPS | Detecting and visualizing hotspots, route planning |
| | NeriCell [26] | Microphone, Accelerometer, GSM Radio, GPS, Camera | Bump, breaking and honking detection |
| Human Behavioral Monitoring | Emotion Sense [27] | Microphone, Accelerometer | Emotion and speaker recognition |
| | Activity Recognition [28] | Accelerometer | Activity recognition |
| Social Networking | CenceMe [29] | Microphone, Camera, GPS, Radio, Accelerometer, etc | Bicycling, Weight lifting, Golf swing analysis |
| | Party Themometer [30] | Microphone, Camera, GPS | Party Themometer |

*2.1. Health Monitoring*

Sha et al. [17] proposed a smartphone assisted chronic illness self-management system (SPA), which aided the prevention and treatment of chronic illness. The SPA can provide continuous monitoring on the health condition of the user and give valuable in-situ context-aware suggestions/feedbacks to improve the public health. To address the growing rate of sedentary lifestyles, Jarvinen et al. [18] developed a system, called UbiFit Garden, which uses small cheaper sensors, real-time statistical modeling, and a personal mobile display to encourage regular physical activity. It resides on the background screen or wallpaper of a mobile phone to provide a subtle reminder whenever and wherever the phone is used. UbiFit Garden relies on the Mobile Sensing Platform (MSP).

*2.2. Environmental Monitoring*

Mun et al. [21] presented the Personal Environmental Impact Report (PEIR) that uses location data sampled from everyday mobile phones to calculate personalized estimates of environmental impact and exposure. It uses mobile handsets to collect and automatically upload data to server-side models that generate web-based output for each participant. One feature that distinguishes PEIR from existing web-based and mobile carbon footprint calculators is its emphasis on how individual transportation choices simultaneously influence both environmental impact and exposure. NoiseTube [22] is a new approach for the assessment of noise pollution involving the general public. The goal of this project is to turn GPS-equipped mobile phones into noise sensors that enable citizens to measure their personal exposure to noise in their everyday environment. Thus each user can contribute by sharing their geo-localized measurements and further personal annotation to produce a collective noise map.

*2.3. Traffic Monitoring*

Thiagarajan et al. [25] proposed a system, called VTrack, for travel time estimation using this sensor data. It addresses two key challenges: energy consumption and sensor unreliability. VTrack performs map matching, which associates each position sample with the most likely point on the road map, and produces travel time estimates for each traversed road segment. VTrack compiles a database of historic travel delays on road segments. Nericell [26] is a system that performs rich sensing by piggybacking on smartphones that users carry with them in normal course. Nericell addresses several challenges including virtually reorienting the accelerometer on a phone that is at an arbitrary orientation, and performing honk detection and localization in an energy efficient manner. The system could be used to annotate traditional traffic maps with information such as the bumpiness of roads, and the noisiness and level of chaos in traffic.

*2.4. Human Behavioral Monitoring*

EmotionSense [27] is a mobile sensing platform for social psychological studies based on mobile phones. The key characteristics of this system include the ability of sensing individual emotions as well as activities, verbal and proximity interactions among members of social groups. This can be used to understand the correlation and the impact of interactions and activities on the emotions and behavior of individuals. Kwapisz et al. [28] described and evaluated a system that uses phone-based accelerometers to perform activity recognition, a task which involves identifying the physical activity a user is performing. The activity recognition model allows to gain useful knowledge about the habits of millions of users passively just by having them carry cell phones in their pockets. Applications of this system includes automatic customization of the mobile devices behavior based upon a user's activity and generating a daily/weekly activity profile to determine if a user is performing a healthy amount of exercise.

## 2.5. Social Networking

Miluzzol et al. [29] presented a system called CenceMe, a personal sensing system that enables members of social networks to share their sensing presence with their buddies in a secure manner. Sensing presence captures a user's status in terms of his activity (e.g., sitting, walking, meeting friends), disposition (e.g. happy, sad, doing OK), habits (e.g. at the gym, coffee shop today, at work) and surroundings (e.g. noisy, hot, bright, high ozone). CenceMe injects sensing presence into popular social networking applications such as Facebook, MySpace, and IM (Skype, Pidgin) allowing for new levels of connection and implicit communication between friends in social networks. Party Thermometer [30] is a human-query application, where queries are directed to users who are at parties. In addition to location, (party) music detection is also employed using the microphone sensor to establish the users' context more precisely.

## 3. Energy Consumption of Smartphones

In order to achieve our purpose of energy saving for smartphones, necessary information about energy consumption of smartphones, such as the composition of energy consumption, the energy consumption of different parts of smartphone, and so forth, is needed. We discuss the energy consumption of smartphones in this section.

### 3.1. Energy Consumption in Operating Mode

In [31], Anand *et al.* evaluated the individual power consumption of the main hardware components (e.g., processor, Wi-Fi interface, backlight) of iMate KJam smartphones shipped with Lithium Polymer 1250mAh battery, 128MB memory, 195 MHz OMAP processor, and Windows Mobile 5.0 operating system. Table 3 illustrates the hardware component distribution of power consumption. As we can see, the processor consumes the most of the power, followed closely by the Wi-Fi interface and the GSM radio. The Bluetooth and backlight cause less power consumption. However, Bluetooth is still a significant cause of power losing because users typically keep it on most of time.

Carroll and Heiser [32] discussed the significance of the power drawn by different components, and analyzed the energy impact of dynamic voltage and frequency scaling of the device's application processor. They depict the power consumed by the display backlight over the range of available brightness levels on Android-based smartphones in operating mode. The minimum backlight power is approximately 7.8 mW, and the maximum is nearly 414 mW.

**Table 3.** Energy consumption for typical use of smartphone.

| Component | Average power consumption percent (%) |
| --- | --- |

| CPU | 35 |
|-----|-----|
| GSM | 25 |
| Wi-Fi | 25 |
| Backlight | 3 |
| Bluetooth | 7 |
| Other | 5 |

### 3.2. Energy Consumption in Sleep

For sake of energy saving, smartphones should power off the radio interface and turn on the radio only when there is on-going traffic. This saves energy for smartphones based on the fact that wireless radios consume much less energy in idle or suspend mode than in active or operating mode. Zhang *et al.* [33] examined the energy consumption of the Wi-Fi radio interface in idle state.

The measurement shows that Wi-Fi requires 1.4260W [34] to scan available networks and requires 0.890W [35] for active transmission. However, it only needs 0.256W in the idle state. The cell radio burns about 2-4mAh in idle state, and costs about 250-300mAh of battery in active state [31]. While Bluetooth consumes 0.01W in idle state, and it costs 0.12W for scanning [13]. We can see that Wi-Fi needs lowest power for data transfer among various network interfaces. However, it consumes significant energy in idle state and incurs a high overhead when scanning. Smartphone costs 268.8mW in the idle state with backlight off, and consumes 68.6mW in the suspended state [32].

### 3.3. Power Consumption of Networks

We conduct a measurement on power efficiency of enhanced data rate for GSM evolution (EDGE) and Wi-Fi radio of ZTE V880 smartphone. As shown in Fig. 1(a), with the increase of bandwidth, sending data requires considerable more energy than receiving. That is, the energy efficiency of receiving is higher than sending in all the cases, and the gap of energy efficiency between receiving and sending is larger when using EDGE to transfer data.

As we can see from Fig. 1(b), with the increase of bandwidth, the energy efficiency of receiving becomes higher than sending in all the cases when Wi-Fi is used to transfer data. However, there is little increase in energy efficiency when the bandwidth is near 50 KB/s, as shown in the Fig. 1(b).

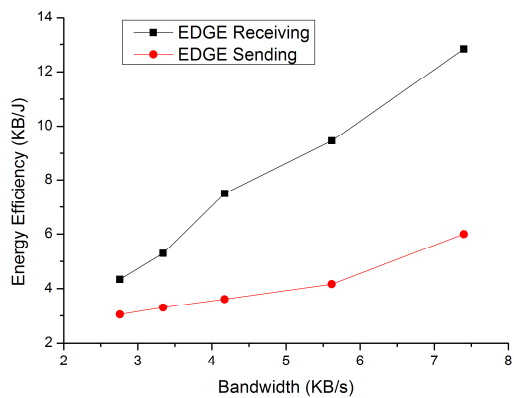

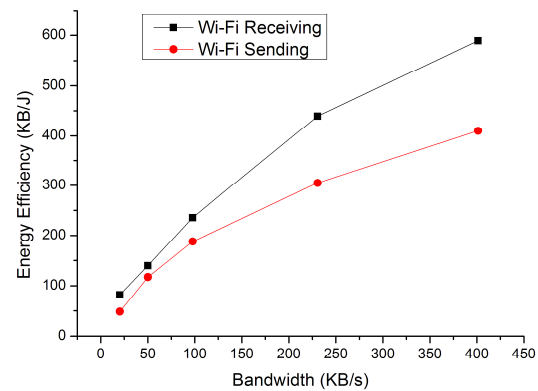

(a)                                                    (b)

**Figure 1.** Consumed energy on ZTE V880 with different bandwidths and methods in use. (**a**) EDGE is in use. (**b**) Wi-Fi is in use.

## 4. State-of-the-art Techniques for Smartphone Energy Saving

A lot of research efforts have been made on reducing energy consumption of smartphones. Depending on the component of smartphone considered, smartphone energy-saving methods can be mainly divided into the following facets.

### 4.1. Battery Modeling

Most smartphones use rechargeable batteries as their mobile power sources, which need to take 1.5 to 4 hours to be fully charged [11]. To use batteries efficiently, we consider them as extremely important resources for the operating system and applications. Battery discharge behavior is affected by various factors, such as discharge rate, temperature [36], and number of charge cycles, etc.

Due to the nonlinearities of battery, it is essential to understand battery discharge behavior, and features for battery designers. Many researches have been done on battery modeling to address this issue. In [37], Fuller *et al.* presented an electrochemical model to describe the charge and discharge of battery. They use concentrated solution theory to model the galvanostatic charge and discharge of a dual lithium ion insertion (rocking-chair) cell. However, it is slow to produce prediction of battery features, and provides limited analytical help for battery designers. Chiasserini and Rao [38] developed a stochastic battery model to represent charge recovery processes. But they just concentrate on charge recovery, and can not account their model for other battery nonlinearities, temperature, and capacity fading, to name a few.

In [39], Manwell and McGowan proposed a kinetic battery model to be used for charging and discharging. They consider the apparent change in capacity as a function of charge and discharge rates. The drawback of this model is that it is not useful for the types of batteries used in mobile computing [40]. Rakhmatov *et al.* [41] presented a diffusion model for battery-aware scheduling algorithm in real-time embedded systems. Briefly, battery designers and programmers should choose different battery models depends on their actual circumstances to better model batteries.

### 4.2. Energy-efficient Communications

Energy-efficient communications have been hot research topic in recent years. Communication components may enter standby or sleep modes to save power. In [33] Zhang *et al.* took advantage of idle state to save the energy consumption of Wi-Fi radio interface. Although this will save the energy consumption, it is hard to estimate the idle state precisely. Shih *et al.* [42] suggested that systems completely turn off the radio interface and turn on the radio only when there is on-going traffic. Nevertheless, this will cause the user completely lose connection to the wireless LAN. In [14] Rozner *et al.* proposed NAPman, a network-assisted power management for Wi-Fi devices, which can leverage AP virtualization and a new energy-aware scheduling algorithm to minimize Wi-Fi devices power consumption.

Bluetooth's energy consumption is very low. Pering *et al.* [35] proposed a scheme that switches between Bluetooth and Wi-Fi interfaces to save battery energy. However, this scheme needs to modify the infrastructure. In [34], Wu *et al.* used the GSM signal to get the location information to estimate the Wi-Fi network AP. The GSM's energy consumption is very small, but its positioning accuracy is unsatisfactory. Blue-Fi [13] uses a combination of Bluetooth contact-patterns and cell-tower information to predict the availability of the Wi-Fi connectivity, and it does not need any modification to the existing infrastructure. However, many mobile devices with Bluetooth radio interfaces change their positions. Bluetooth has a much lower range compared to other network technologies. Therefore, mobile devices with Bluetooth can not be used for our problem.

In [15] Zhou *et al.* utilized ZigBee radios to identify the existence of Wi-Fi networks through unique interference signatures generated by Wi-Fi beacons. They incur extra fees to buy or employ ZigBee cards, even though they detect Wi-Fi network APs with high accuracy, short delay and little computation overhead.

There is an idea of using a separate low-powered radio to wake up a high-powered radio. Like [42] Wake-on-Wireless uses a low power radio to serve as a wake-up channel for a Wi-Fi radio. However, it needs significant modifications to existing mobile devices. While On-Demand-Paging [43] builds on this idea to use the widely available Bluetooth radios as the low-powered channel, it also requires substantial infrastructure support in the form of specialized APs that have both Wi-Fi and Bluetooth interfaces. Cell2Notify [44] uses the cellular interface to wake up the Wi-Fi interfaces on an incoming VOIP call using specialized servers.

*4.3. Energy-efficient Computation*

We describe two basic approaches, including putting processor into sleep mode and computation offloading, to save CPU's energy in this section.

Brakmo *et al.* [45] present an energy reduction technique, called uSleep, for handheld devices, which is most effective when the device's processor is lightly loaded. They try to put the processor in sleep mode rather than in idle mode to save energy. However, it is not effective when the device's processor is highly loaded.

With computation offloading, the mobile device does not perform the computation; instead, computation is performed somewhere else, such as on a server, thereby extending battery lifetime. In [46], Pathak *et al.* designed and implemented an event-tracing-based profiling tool, called XRay. The tool identifies methods of an application that can be offloaded to a remote server, and determines whether and when the methods are offloaded to server will save energy. Needing a priori knowledge of input parameters and network conditions is the drawback of the tool. Xian *et al.* [47] performed offloading at the level of applications. They execute the program initially on the mobile client with a timeout, and offload the program to the server when it is not completed after the timeout. However, many applications that we can not offload entirely are just fit for offloading parts of applications to server. Kumar and Lu [48] applied computation offloading from a remote server to cloud platform. It is new to apply computation offloading in cloud, and more exploration needs to be done.

*4.4. Energy-Efficient Mobile Sensing*

Mobile device based sensing can provide rich contextual information for mobile applications such as social networking and location-based services. However, the sensors on mobile devices consume too much energy. It will limit the continuous functioning of the mobile applications on mobile devices because of these devices' limited battery capacity. Consequently, energy-efficient mobile sensing [2, 16, 49-51] has become a hot topic in mobile computing.

Abdesslem et al. [49] proposed to use less energy-consumption sensors more often instead of more energy-consumption sensors. By choosing when to use more energy-efficient sensors, it is possible to reduce the energy consumption of mobile sensing applications. Constandache et al. [16] proposed a location sensing adaptive framework called EnLoc. The framework characterizes the optimal localization accuracy for a given energy budget and developed prediction-based heuristics for real-time use. By taking into account the accuracy-energy trade-off of different location sensors available in mobile phones, EnLoc selects the energy-optimal sensor and reduces energy consumption. Kang et al. [50] proposed a scalable and energy-efficient monitoring framework called SeeMon for sensor-rich and resource-limited mobile environments. The authors explored the hierarchical sensor management concept which achieves energy efficiency by only performing context recognition when changes occur during the context monitoring. Priyantha et al. [51] designed a novel system architecture called Little Rock at the hardware level for energy-efficient sensing. They proposed to add a low power microcontroller or an additional low power core in the multi-core processor responsible for managing sensors. This kind of modification enables phones to enter a low power sleep state while the low power sensor processor is continuously sampling and processing sensor data. Wang et al. [2] presented a novel design framework for an Energy-Efficient Mobile Sensing System (EEMSS). EEMSS uses hierarchical sensor management strategy to recognize user states as well as to detect state transitions. By powering only a minimum set of sensors and using appropriate sensor duty cycles the device energy consumption can be significantly reduced.

Energy-efficient mobile sensing has been studied for a long time but there are still some unresolved issues. Activity recognition needs more effective matching algorithm to improve the recognition accuracy. With the increasing number of sensors, how to manage them in an energy-efficient manner is a problem. In addition, privacy should be paid much attention to because lots of user information in mobile sensing system might be used illegally.

*4.5. Other Energy Saving Approaches*

There are many research efforts [52-54] focusing on reducing the energy of display and memory. The code involving graphical user interface (GUI) usually accounts for 48 percent of the total code of an application [55]. As GUI always invokes display to show what the application does, display also consumes a large amount of energy. Zhong and Jha optimized GUIs for energy consumption in [52]. They present the first GUI energy characterization methodology, improve GUI platforms, and design GUIs in an energy-efficient way. However, the power consumption of display is highly depends on users' activity patterns that are hard to predict. Therefore, it's difficult to save display's energy.

Aho et al. [54] reduced the number of memory operands to save memory's energy. They optimally allocate temporaries and globally register for the most frequently used variables during compile phase. Although this approach can save energy, it incurs the overhead of longer compilation time and compiler complexity, especially for commercial software.

## 5. Location-assisted Wi-Fi Discovery

### 5.1. System Overview

This section presents the key components and operation workflow of the proposed solution to improve the energy efficiency of smartphones, see Fig. 2. Our system includes three modules: rate monitor module, switching decision engine module and switching module. On the mobile devices, we use rate monitor module to measure cellular data rates (both uplink and downlink) at a chosen interval $T_{measure}$ (see Table 4) periodically. Based on the collected information, the switching decision engine module decides whether the users need to switch to Wi-Fi network: if yes, then it invokes the switching module to discover the nearest Wi-Fi network AP, and users switch to Wi-Fi network when they arrive at it. Below we will describe the three modules in detail.

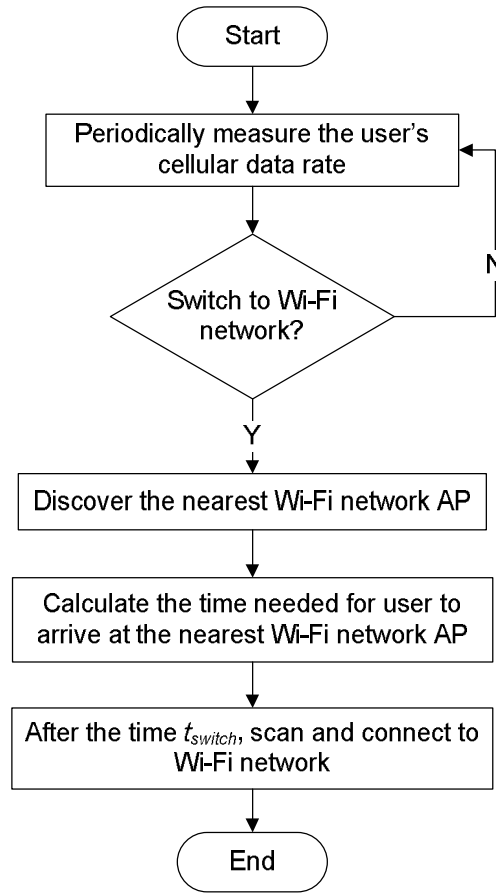

**Figure 2.** Workflow of the system.

**Table 4.** List of notations.

| Variable | Description |
| --- | --- |
| $T_{measure}$ | The data rate collect interval |
| $R_{user}$ | The data rate of the user |
| $B_t$ | The threshold of bandwidth that is set by the user |
| $P_{near}$ | The nearby Wi-Fi network AP's location |
| $P_{user}$ | The user's location |
| $Lat_{user}$ | The latitude of the user's location |

| $Lon_{user}$ | The longitude of the user's location |
|---|---|
| $P_{ap}$ | The nearest Wi-Fi network AP's location |
| $Lat_{ap}$ | The latitude of the nearest Wi-Fi network AP's location |
| $Lon_{ap}$ | The longitude of the nearest Wi-Fi network AP's location |
| $d$ | The distance between user and the nearest Wi-Fi network AP |
| $R$ | The radius of the earth |
| $v_{user}$ | The speed of the user |
| $t_{switch}$ | The time needed for user to go from his/her location to the nearest Wi-Fi network AP |

### 5.1.1. Rate Monitor Module

To adapt to condition changes, our system must monitor the conditions of the mobile device and the wireless network. In [56] Yan *et al.* used available bandwidth, packet loss, received signal strength, energy consumption, operator requirements, user preferences, etc. to guide the selection of the best network. In [57] Chamodrakas and Martakos proposed a novel network selection based on TOPSIS [58] that takes into account both network conditions and user preferences as well as QoS and energy consumption to select the best network. In this paper, in order to simplify the situation and reduce unnecessary cost in those points, we just use the data rate of the user to guide switching decision making. Whenever some significant change happens (for example, a large data rate of the user fluctuation occurs), the switching decision engine decides whether to trigger switching. This rate monitor module is designed to monitor the data rate of users, which is used to guide switching decision making for our system.

### 5.1.2. Switching Decision Engine Module

To perform switching decision engine, our system examines the data collected by rate monitor module. It then decides whether to trigger switching module according to user's switching goals. If so, it decides what level of $R_{user}$ to use on mobile device, that is, how much bandwidth the user needs. We use a simple threshold-based approach to decide whether to trigger switching module. If the $R_{user}$ is more than or equal to the threshold $B_t$, then the switching decision engine module triggers switching module and switches to a wide bandwidth Wi-Fi network. To adapt to the user preferences, the user can set different thresholds and chooses the appropriate threshold.

### 5.1.3. Switching Module

In this module, our system decides when and how to switch to Wi-Fi networks. Fig. 3 shows the workflow of switching module of our proposed system, and it is discussed below.

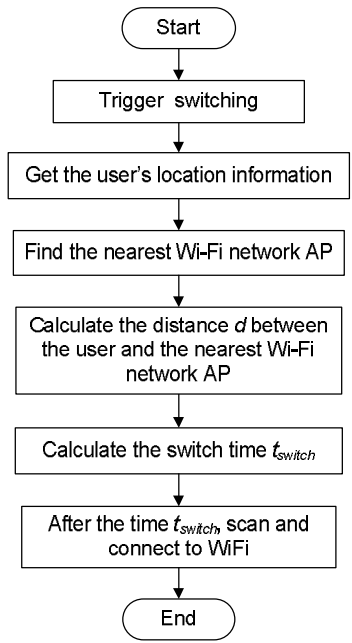

**Figure 3.** Workflow of switching module.

When the switching module is triggered, it takes four steps to complete the switching process. In the first step, it uses location sensors to get the user's location. We use different location sensors in different environments. As we have already mentioned, considering energy-accuracy tradeoff and restrictions of different location sensors, we adopt A-GPS in outdoors and ZigBee in indoors to get user's location information. Thus the switching module consists of two submodules, that is, the outdoor switching submodule and the indoor switching submodule.

In the second step, it finds the nearest Wi-Fi network AP's location $P_{ap}$ by calculating the distance between the user's location $P_{user}$ and the nearby Wi-Fi network AP's location $P_{near}$. In the third step, it calculates the distance $d$ between the user and the nearest Wi-Fi network AP.

In the last step, we assume that the user moves towards the nearest Wi-Fi network AP at a constant speed $v_{user}$ and there are no obstacles between the user and the nearest Wi-Fi network AP. We then use (1) to calculate the time needed for the user to go from his/her location to the nearest Wi-Fi network AP. After the time $t_{switch}$, the user begins to scan and connect to the Wi-Fi network.

$$t_{switch} = d/v_{user} \tag{1}$$

### 5.2. Outdoor Switching Submodule

Here we describe an *A-GPS assisted network switching scheme* (*A-GPS switching* for short) in outdoors. We will compare the power consumption of this scheme with other four schemes, which will be discussed later.

The operational workflow of this scheme is shown in Fig. 4. Our system uses A-GPS to locate the user in outdoors. Then it discovers the nearest Wi-Fi network AP by calculating the distance between the user's location and the nearby Wi-Fi network AP's location and gets its location. We can calculate the distance between user and nearest Wi-Fi network AP by using (2), and use (1) to calculate the time needed to switch to Wi-Fi network. In (2), $Lat_{user}$, $Lon_{user}$, $Lat_{ap}$ and $Lon_{ap}$ are expressed in radians, and the units of $d$ and $R$ are kilometer. Finally, we start to scan and connect to the nearest Wi-Fi network AP after the time $t_{switch}$.

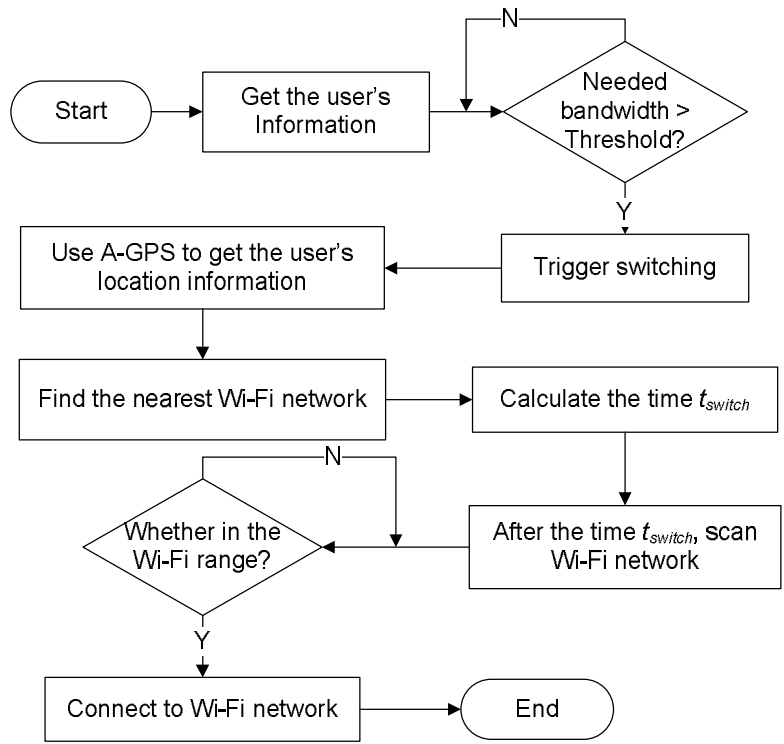

**Figure 4.** Workflow of A-GPS assisted network switching scheme.

We assume that we already know the distribution of the Wi-Fi network around the user, and this distribution is relatively stable. In a nutshell, we discover the Wi-Fi network by periodically scanning its signal. When a mobile device discovers a Wi-Fi network, it logs this Wi-Fi network in a log $L$, locally. The log entries are in the form of (*Timestamp*, {*Wi-Fi network*}, *Location*, *Range*). Wi-Fi network is identified by its SSID/BSSID, and location is identified by Wi-Fi network AP's location which is got by the Wi-Fi sensors. Range represents how far the Wi-Fi network signals reach. Given a user's location, we check the log whether the user is in the range of a Wi-Fi network: if yes, then we consider the user can connect to the Wi-Fi network.

$$d = 2\arcsin\sqrt{\sin^2\frac{Lat_{user} - Lat_{ap}}{2} + \cos Lat_{user} \times \cos Lat_{ap} \times \sin^2\frac{Lon_{user} - Lon_{ap}}{2}} \times R \qquad (2)$$

### 5.3. Indoor Switching Submodule

A *ZigBee assisted network switching scheme* (*ZigBee switching* for short) is presented in this submodule for indoors. Fig. 5 shows the workflow of this scheme. If switching module is triggered, our system adopts ZigBee to get the user's location in indoors. Then it calculates the distance between the user's location and the nearby Wi-Fi network AP's location to find the nearest Wi-Fi network AP's location. Thus it can get the distance between user and nearest Wi-Fi network AP. Finally, it calculates the time needed for user to go from his/her location to the nearest Wi-Fi network AP by using (1). After that, the user begins to scan and connect to the Wi-Fi network.

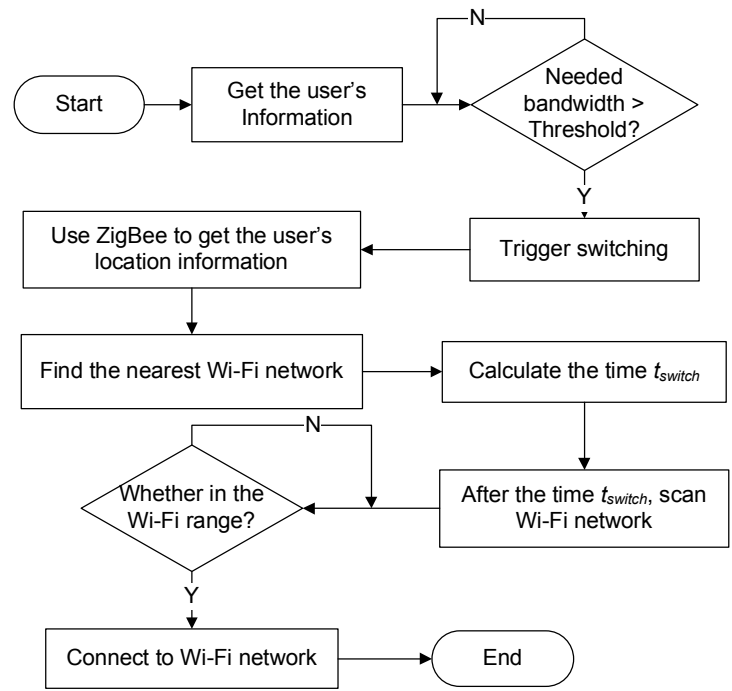

**Figure 5.** Workflow of ZigBee assisted network switching scheme.

## 6. Experiments for Outdoor Scenarios

In this section, we use the log $L$ to evaluate five outdoor schemes with appropriate thresholds and energy efficiency for different users.

### 6.1. Experimental Setup

The experiments are based on a scenario that a user lives in a suburban school area, as shown in Fig. 6, where each circle represents the approximate coverage area of each network. We use ZTE X876 smartphone for our data collection. It is a Google Android 2.2 smartphone with integrated Wi-Fi interface and is capable of EDGE data connectivity. It has a battery capacity of 1500mAh at 3.7 volts. We have implemented our system, called *SwitchR*. The red circle represents a Wi-Fi network and the blue one represents the user. We also developed other three software tools, which will be discussed in the following three parts, to record our experimental data with minimal intrusion to the normal smartphone operation.

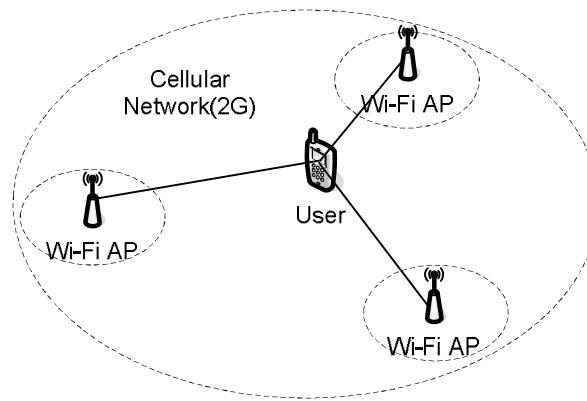

**Figure 6.** Networks used in experiments.

### 6.1.1 Log Collection

Our first software, called *Log Collector*, records the Wi-Fi network around the user. We discover the Wi-Fi network by periodically scanning its signal. We go around the suburban school area, and when we discover a Wi-Fi network, the *Log Collector* logs this Wi-Fi network in the log $L$, locally. Therefore, given a user's location, we can easily know whether the user is in the range of a Wi-Fi network by checking the log $L$.

### 6.1.2 Rate Collection

We use the second software, called *Rate Collector*, to measure different users' data rates every 30 seconds, as shown in Fig. 7(a). As we can see from the figure, the data rate of the user in a continuous 60 minutes experiment jitters seriously and changes from 0KB/s to 60KB/s. When the experiment starts, it begins to record user's data rate periodically. The first version of *Rate Collector* records user's data rate every 10 seconds, which is too frequent to represent the user's average data rate.

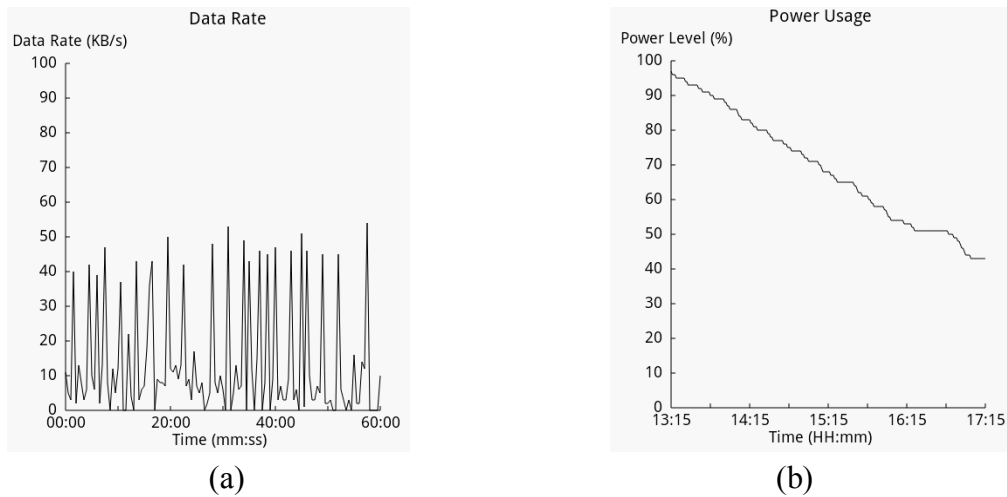

(a)                                                                (b)

**Figure 7. (a)** Data rate of the user in a continuous 60 minutes. **(b)** Remaining capacity of battery in a 4-hour time period experiment.

### 6.1.3 Measuring Phone Power Consumption

We have developed software based on our previous work [59], called *PowerUsage*, to measure the power consumption of our smartphone under controlled conditions. We use battery interface provided by Google APIs to record the current percent of the whole battery every minute. We carried out an experiment to use it to get the data from a 4-hour time period between 13:15 and 17:15. Fig. 7(b) shows the remaining capacity of battery changes from 97 percent to 43 percent in the 4-hour time period experiment.

### *6.2. Schemes for Comparison*

Here we describe other four experimental schemes, which are compared to the A-GPS switching scheme in this section. Two switching schemes are considered. In the first scheme, called *GSM assisted network switching scheme* (*GSM switching* for short), we use GSM to get the user's location.

Then we get the nearest Wi-Fi network AP in the log *L* by using the user's location information. Therefore, we can calculate the distance between user and nearest Wi-Fi network AP, and the time needed to switch to Wi-Fi network. In the second scheme, called *scanning assisted network switching scheme* (*Scanning switching* for short), we do not use the user's location information. When the switching module is triggered, the user scans for Wi-Fi networks. The user connects to Wi-Fi network when he or she discovers one. We assume that the user moves towards the nearest Wi-Fi network AP, that is, we measure the lower bound of power consumption.

In addition, two non-switching schemes will also be examined. In the first (non-switching) scheme, called *always use GPRS* (*GPRS non-switching* for short), we do not switch and always use GPRS to access the internet. While in the second scheme, called *always use Wi-Fi* (*Wi-Fi non-switching* for short), we always use Wi-Fi to surf the internet and do not switch as well. Moreover, in these two non-switching schemes, GPRS and Wi-Fi are always available for us, and there is no need to switch between networks.

In order to compare the five schemes, we divide users into four categories according to their behaviors (see Table 5). We will evaluate the five schemes in four thresholds (see Table 6) for different users.

**Table 5.** Various user characteristics.

| User | Application(s) | Bandwidth |
|------|----------------|-----------|
| U1 | Text message | Low |
| U2 | Text message and Web browsing | Medium |
| U3 | Text message and Video streaming | High |
| U4 | Text message and File downloading | Very High |

**Table 6.** Different thresholds.

| Threshold | Value |
|-----------|-------|
| B1 | 5 kb/s |
| B2 | 10 kb/s |
| B3 | 15 kb/s |
| B4 | 20 kb/s |

*6.3. Energy Consumption*

We evaluate the energy consumption of five outdoor schemes for different users under different thresholds. Using the full battery capacity of the smartphone, we measure the power consumption of different schemes before the phone runs out of power (total capacity of 19980 J).

Fig. 8(a) shows the energy efficiency of five outdoor schemes for the user U1 under different thresholds $B_1$, $B_2$, $B_3$ and $B_4$. We observe that our *A-GPS switching scheme* is close to other two switching schemes. Ideally, they are the same in terms of energy efficiency due to the user U1 does not need switch from GPRS to Wi-Fi network. Our scheme consumes a little more power than the best scheme of five outdoor schemes because of the overhead of our system itself and high power efficiency for data transfers of the Wi-Fi network. As we can see from Fig. 8(a), the user U1 does not need high bandwidth. Thus the bandwidth of GPRS is enough for U1, and the user U1 does not need any threshold.

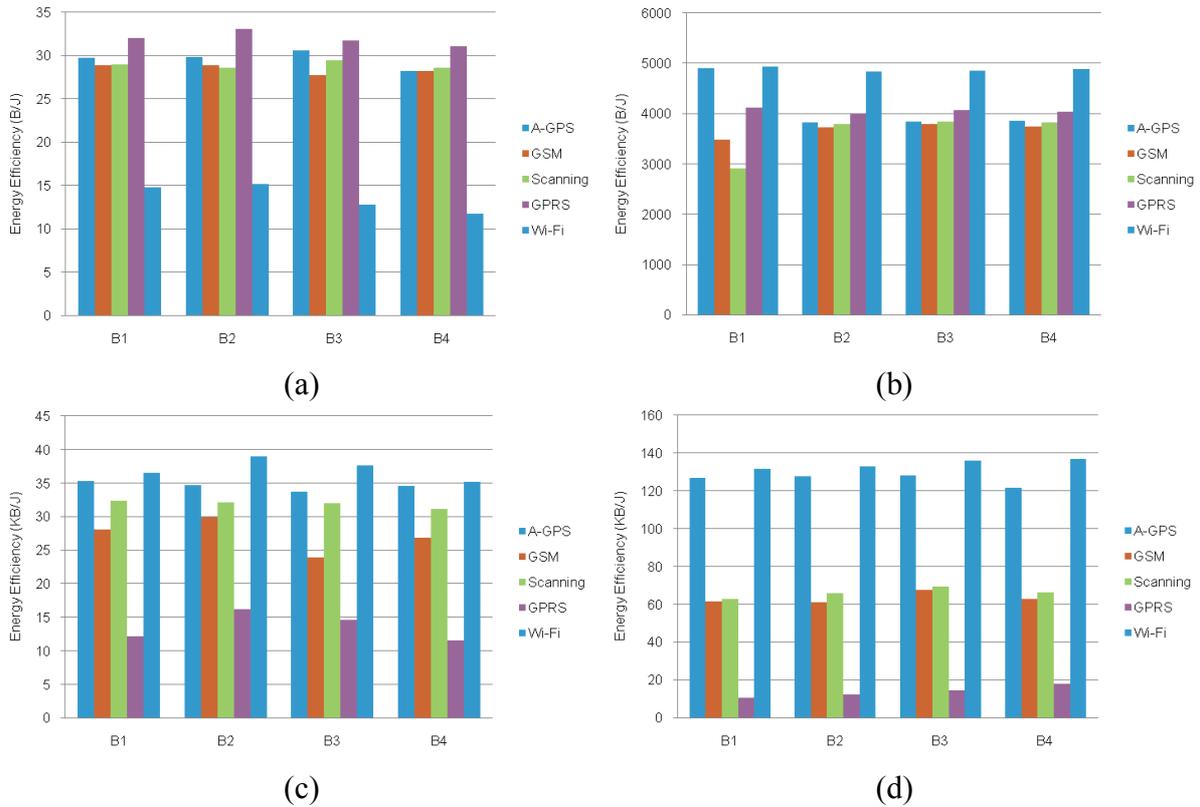

**Figure 8.** Energy efficiency in experiments. (**a**) U1. (**b**) U2. (**c**)U3. (**d**) U4.

We plotted the energy efficiency of five outdoor schemes for the user U2 under different thresholds $B_1$, $B_2$, $B_3$ and $B_4$ in Fig. 8(b). As we can see from the figure, our scheme is the best among the three switching schemes under the threshold $B_1$. There are two reasons behind this. First, A-GPS provides arguably the best combination of energy and accuracy for location sensing compare with GPS and GSM. Second, our scheme greatly reduces the numbers of unnecessary Wi-Fi scans on the mobile devices compare with other switching schemes. We observe that the user U2 does not need switch to Wi-Fi network under the thresholds $B_2$, $B_3$ and $B_4$. Therefore, the three switching schemes are close, and the appropriate threshold for the user U2 is the threshold $B_1$.

As shown in Fig. 8(c), our scheme is always the best among the three switching schemes. The reasons behind this have been discussed previously. As we can see from Fig. 8(c), the user U3 always switches from GPRS to Wi-Fi network under the different thresholds. Our scheme also consumes a little more energy than the best scheme of five schemes due to the overhead of our system itself and A-GPS. Fig. 8(c) shows the user U3 needs high bandwidth, and the appropriate threshold for the user U3 is the threshold $B_4$.

The user U4 downloads a 50MB file from the Internet. Fig. 8(d) shows the energy efficiency of five schemes for the user U4 under different thresholds $B_1$, $B_2$, $B_3$ and $B_4$. In our experiments, our scheme is always the best of three switching schemes. Because of many unnecessary Wi-Fi scans of the *scanning switching scheme*, it wastes too much energy. Due to the high inaccuracy of GSM positioning, the *GSM switching scheme* also does many unnecessary Wi-Fi scans on the mobile devices. This leads to its bad performance. Our scheme also consumes a little more energy than the best scheme, i.e. the *Wi-Fi non-switching scheme*. As we can see from Fig. 8(d), the user U4 needs very high bandwidth and the appropriate threshold for the user U4 is also the threshold $B_4$.

As shown in Fig. 9, the GPRS has a better performance than the Wi-Fi network under the user U1. However, under the users U2, U3 and U4, the Wi-Fi network has a much better performance than the GPRS. We observe that there is an approximate exponential increase in the average energy efficiency ratio of Wi-Fi to GPRS. It can be concluded that Wi-Fi provides the best combination of bandwidth and power efficiency for data transfers in the examined scenarios.

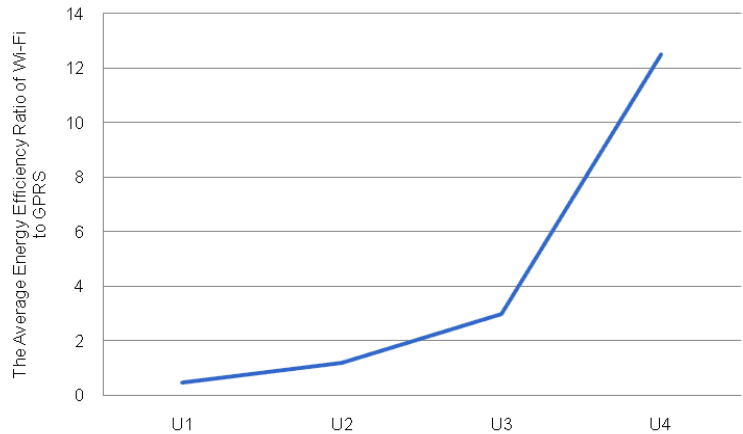

**Figure 9.** Average energy efficiency ratio of Wi-Fi to GPRS.

## 7. Simulations for Indoor Scenarios

We use OMNeT++ to evaluate the five indoor schemes with appropriate thresholds and energy efficiency for different users in this section.

### 7.1. Simulation Setup

The simulations are based on a scenario that a user lives or works in a building. We deploy several ZigBee nodes, which are used to locate the user, in the building, and its approximate positioning accuracy is set to 5 meters. We also randomly deploy forty Wi-Fi network APs in the rectangle, of which the length is 2000 meters and width is 340 meters, around the building. The coverage area of each network is 100 meters. We assume that the user goes to the nearest Wi-Fi network AP at a constant speed of 1m/s, when he or she finds one. We also use the ZTE X876 smartphone as the mobile client for our simulations.

We will compare other four schemes, which are introduced in Section 6.2, to the *ZigBee switching scheme* and evaluate the energy consumption of five indoor schemes in four thresholds (see Table 6) for different users (see Table 5).

### 7.2. Energy Consumption

We compare the energy consumption of five indoor schemes for different users under different thresholds. The results of simulations are shown in Fig.10. As we can see from the figure, the results of indoor simulations with different bandwidths under different schemes are similar to that of outdoor experiments. Consequently, we skip here the detailed analysis for the sake of simplicity.

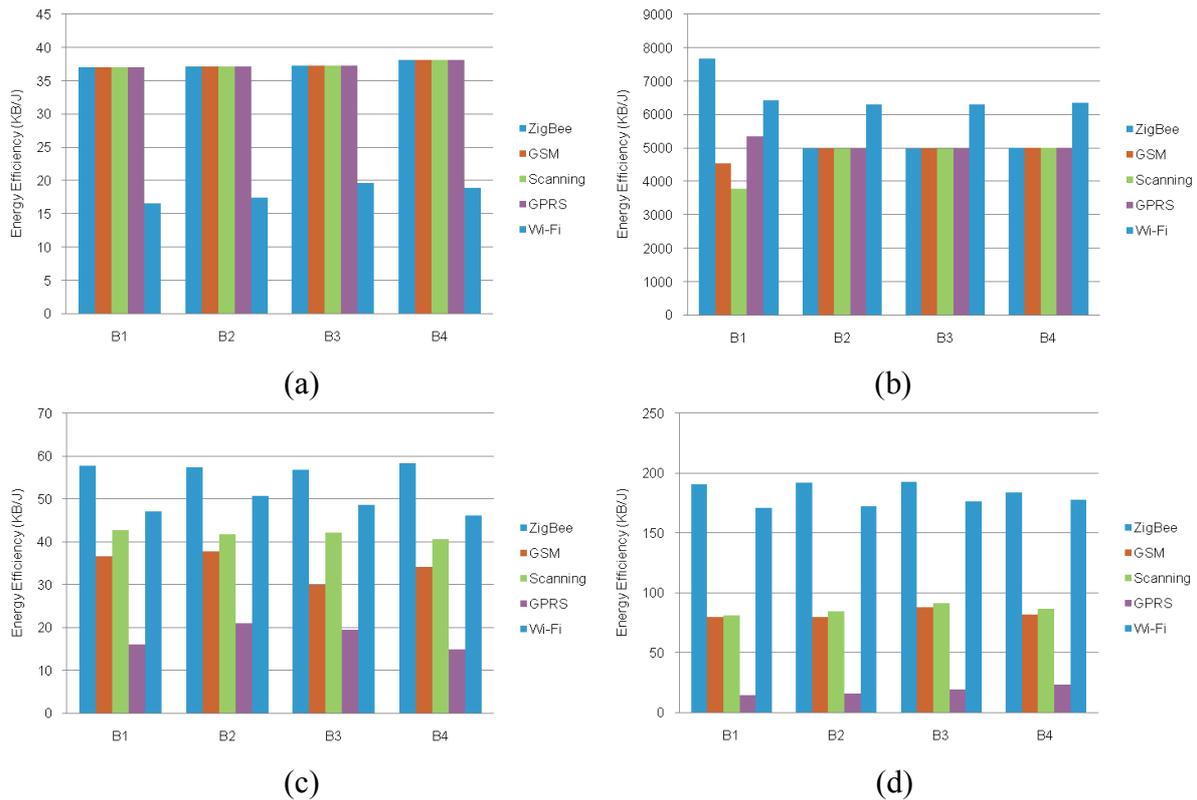

**Figure 10.** Energy efficiency in simulations. (**a**) U1. (**b**) U2. (**c**)U3. (**d**) U4.

## 8. Conclusions

In this paper we have reviewed some typical smartphone applications, and discussed the power consumption of smartphones. We have conducted a measurement on power efficiency of EDGE and Wi-Fi radio of ZTE V880 smartphone. We have also summarized basic techniques of energy saving for smartphones. Furthermore, we have proposed a location-assisted Wi-Fi discovery scheme that uses user's location information to discover the nearest Wi-Fi network AP. We leveraged complementary characteristics of Wi-Fi and GPRS to meet the user bandwidth needs and minimize the power consumption of smartphones. The proposed scheme allows the user to switch to the Wi-Fi interface intelligently when he or she arrives at the nearest Wi-Fi network AP. Thus it considerably reduces the number of unnecessary Wi-Fi scans on non-connected state. Our experiments and simulations show that it effectively saves energy for smartphones in certain scenarios.

However, we simply used the data rate of the user to make switching decision. We did not take other factors, such as network conditions and QoS, into account to select the best network. Therefore, our switching decision making procedure is not flexible enough to handle various conditions. As we can see from our experiments, different users have different appropriate thresholds. Thus our system must be adaptable to different users. Further research is necessary in this regard. In addition, we will upload every mobile device's log $L$ and offload the matching computations to the cloud computing platform in future work.

**Acknowledgments**

This work was partially supported by Liaoning Provincial Natural Science Foundation of China under Grant No.201202032, the Fundamental Research Funds for Central Universities (DUT12JR10), and the Innovation Fund of School of Software, Dalian University of Technology.

## References

1. Iftode, L.; Borcea, C.; Ravi, N.; Kang, P.; Zhou, P. Smart Phone: an embedded system for universal interactions. In Proceedings of the 10th IEEE International Workshop on Future Trends of Distributed Computing Systems, Suzhou, China, May 2004; pp. 88-94.

2. Y. Wang,; Lin, J.; Annavaram, M.; Jacobson, Q. Q.; Hong, J.; Krishnamachari, B.; Sadeh, N. A Framework of Energy-Efficient Mobile Sensing for Automatic User State Recognition. In Proceedings of the 7th International Conference on Mobile Systems, Applications, and Services, Kraków, Poland, June 2009; pp. 179-192.

3. Horanont, T.; Shibasaki, R. An implementation of mobile sensing for large-scale urban monitoring. In Proceedings of International Workshop on Urban, Community, and Social Applications of Networked Sensing Systems, Raleigh, NC, USA, November 2008; pp. 51-55.

4. Ishida, Y.; Konomi, S.; Thepvilojanapong, N.; Suzuki, R.; Sezaki, K.; Tobe, Y. An Implicit and User-Modifiable Urban Sensing Environment. In Proceedings of International Workshop on Urban, Community, and Social Applications of Networked Sensing Systems, Raleigh, NC, USA, November 2008; pp. 36-40.

5. O'Hara, K.; Kindberg, T.; Glancy, M.; Baptista, L.; Sukumaran, B.; Kahana, G.; Rowbotham, J. Collecting and sharing location-based content on mobile phones in a zoo visitor experience. *CSCW* **2007**, *16*, 11-44.

6. Michel, S.; Salehi, A.; Luo, L.; Dawes, N.; Aberer, K.; Barrenetxea, G.; Bavay, M.; Kansal, A.; Kumar, K. A.; Nath, S.; *et al.* Environmental Monitoring 2.0. In Proceedings of the 25th International Conference on Data Engineering, Shanghai, China, March-April, 2009; pp. 1507-1510.

7. Hoh, B.; Gruteserand, M.; Herring, R.; Ban, J.; Work, D.; Herrera, J.; Bayen, A. M.; Annavaram, M.; Jacobson, Q. Virtual trip lines for distributed privacy-preserving traffic monitoring. In Proceedings of 6th Annual International Conference on Mobile Systems, Applications and Services, Breckenridge, CO, USA, June 2008; pp. 15-28.

8. Jones, V.; Gay, V.; Leijdekkers, P. Body Sensor Networks for Mobile Health Monitoring: Experience in Europe and Australia. In Proceedings of the 4th International Conference on Digital Society, St. Maarten, Netherlands Antilles, February 2010; pp. 204-209.

9. Taylor, I. M.; Labrador, M. A. Improving the energy consumption in mobile phones by filtering noisy GPS fixes with modified Kalman filters. 2011 IEEE Wireless Communications and Networking Conference (WCNC). Quantana-Roo, Mexico, March 2011; pp. 2006-2011.

10. Xiao, Y.; Bhaumik, R.; Yang, Z.; Siekkinen, M.; Savolainen, P.; Ylä-Jääski, A. A system-level model for runtime power estimation on mobile devices. 2010 IEEE/ACM International Conference on Green Computing and Communications (GreenCom) & 2010 IEEE/ACM International Conference on Cyber, Physical and Social Computing (CPSCom), Hangzhou, China, December 2010; pp. 27-34.

11. Rao, R.; Vrudhula, S.; Rakhmatov, D. N. Battery modeling for energy aware system design. *Computer* **2003**, *36*, 77-87.

12. Rahmati, A.; Zhong, L. Context-for-wireless: context-sensitive energy-efficient wireless data transfer. In Proceedings of The 5th International Conference on Mobile Systems, Applications, and Services, San Juan, Puerto Rico, June 2007; pp. 165-178.

13. Ananthanarayanan G.; Stoica, I. Blue-Fi: enhancing Wi-Fi performance using bluetooth signals. In Proceedings of The 7th International Conference on Mobile Systems, Applications, and Services, Kraków, Poland, June 2009; pp. 249-261.

14. Rozner, E.; Navda, V. NAPman: network-assisted power management for WiFi devices. In Proceedings of the 8th International Conference on Mobile Systems, Applications, and Services (MobiSys '10), San Francisco, CA, USA, June 2010; pp. 91-105.

15. Zhou, R.; Xiong, Y.; Xing, G.; Sun, L.; Ma, J. ZiFi: wireless LAN discovery via ZigBee interference signatures. In Proceedings of the 16th Annual International Conference on Mobile Computing and Networking (MobiCom '10), Chicago, IL, USA, September 2010; pp. 49-60.

16. Constandache, I.; Gaonkar, S.; Sayler, M.; Choudhury, R. R.; Cox, L. EnLoc: energy-efficient localization for mobile phones. INFOCOM, Rio de Jaeiro, Brazil, April 2009; pp. 2716-2720.

17. Sha, K.; Zhan, G.; Shi, W.; Lumley, M.; Wiholm, C.; Arnetz, B. Spa: a smartphone assisted chronic illness self-management system with participatory sensing. In Proceedings of the 2nd International Workshop on Systems and Networking Support for Health Care and Assisted Living Environments, Breckenridge, CO, USA, June 2008; 5:1-5:3.

18. Jarvinen, P.; Jarvinen, T. H.; Lahteenmaki, L.; Sodergard, C. HyperFit: hybrid media in personal nutrition and exercise management. In Proceedings of 2nd International Conference on Pervasive Computing Technologies for Healthcare, Tampere, Finland, January-February 2008; pp.222-226.

19. Denning, T.; Andrew, A.; Chaudhri, R.; Hartung, C.; Lester, J.; Borriello, G.; Duncan, G. Balance: towards a usable pervasive wellness application with accurate activity inference. In Proceedings of the 10th workshop on Mobile Computing Systems and Applications, Santa Cruz, CA, USA, February 2009; 5.

20. Sashima, A.; Inoue, Y.; Ikeda, T.; Yamashita, T.; Kurumatani, K. Consorts-s: a mobile sensing platform for context-aware services. In Proceedings of International Conference on Intelligent Sensors, Sensor Networks and Information 2008, Sydney, Australia, December 2008; pp. 417-422.

21. Mun, M.; Reddy, S.; Shilton, K.; Yau, N.; Burke, J.; Estrin, D.; Hansen, M.; Howard, E.; West, R.; Boda, P. Peir, the personal environmental impact report, as a platform for participatory sensing systems research. In Proceedings of the 7th International Conference on Mobile Systems, Applications, and Services, Kraków, Poland, June 2009; pp. 55-68.

22. Maisonneuve, N.; Stevens, M.; Niessen, M. E.; Steels, L. Noisetube: Measuring and mapping noise pollution with mobile phones. *Information Technologies in Environmental Engineering* **2009**, *2*, 215-228.

23. Kanjo, E.; Benford, S.; Paxton, M.; Chamberlain, A.; Fraser, D. S,; Woodgate, D.; Crellin, D.; Woolard, A. MobGeoSen: facilitating personal geosensor data collection and visualization using mobile phones. *Personal and Ubiquitous Computing* **2008**, *12*, 599-607.

24. Bilandzic, M.; Banholzer, M.; Peev, D.; Georgiev, V.; Balagtas-Fernandez, F.; De Luca, A. Laermometer: a mobile noise mapping application. In Proceedings of the 5th Nordic conference on Human-computer interaction: building bridges, Lund, Sweden, October 2008; pp. 415-418.

25. Thiagarajan, A.; Ravindranath, L.; LaCurts, K.; Madden, S.; Balakrishnan, H. Toledo, S.; Eriksson, J. Vtrack: accurate, energy-aware road traffic delay estimation using mobile phones. In Proceedings of the 7th ACM Conference on Embedded Networked Sensor Systems, Berkeley, CA, USA, November 2009; pp. 85-98.

26. Mohan, P.; Padmanabhan, V. N.; Ramjee, R. Nericell: rich monitoring of road and traffic conditions using mobile smartphones. In Proceedings of the 6th ACM conference on Embedded network sensor systems, Raleigh, NC, USA, November 2008; pp. 323-336.

27. Rachuri, K. K.; Musolesi, M.; Mascolo, C.; Rentfrow, P. J.; Longworth, C.; Aucinas, A. Emotionsense: a mobile phones based adaptive platform for experimental social psychology research. In Proceedings of the 12th ACM International Conference on Ubiquitous Computing. Copenhagen, Denmark, September 2010; pp. 281-290.

28. Kwapisz, J. R.; Weiss, G. M.; Moore, S. A. Activity recognition using cell phone accelerometers. *ACM SIGKDD Explorations Newsletter* **2010**, *12*, 74-82.

29. Miluzzo, E.; Lane, N. D.; Eisenman, S. B.; Campbell, A. T. Cenceme: injecting sensing presence into social networking applications. *Smart Sensing and Context* **2007**, *4793*, 1-28.

30. Das, T.; Mohan, P.; Padmanabhan, V. N.; Ramjee, R.; Sharma, A. Prism: platform for remote sensing using smartphones. In Proceedings of the 8th International Conference on Mobile Systems, Applications, and Services, San Francisco, CA, USA, June 2010; pp. 63-76.

31. Anand, A.; Manikopoulos, C.; Jones, Q.; Borcea, C. A quantitative analysis of power consumption for location-aware applications on smart phones. In Proceedings of IEEE International Symposium on Industrial Electronics, Vigo, Spain, June 2007; pp. 1986-1991.

32. Carroll, A.; Heiser, G. An analysis of power consumption in a smartphone. In Proceedings of the 2010 USENIX Annual Technical Conference, Boston, MA, USA, June 2010; p. 21.

33. Zhang, T.; Madhani, S.; Gurung, P.; van den Berg, E. Reducing energy consumption on mobile devices with WiFi interfaces. In Proc. Global Telecommunications Conf. (Globecom 05), St. Louis, MO, USA, November-December 2005; pp. 561–565.

34. Wu, H.; Tan, K.; Liu, J.; Zhang, Y. Footprint: cellular assisted Wi-Fi AP discovery on mobile phones for energy saving, WiNTECH'09 (MobiCom workshop), Beijing, China, September 2009; pp. 67-75.

35. Pering, T.; Agarwal, Y.; Gupta, R.; Want, R. Coolspots: reducing the power consumption of wireless mobile devices with multiple radio interfaces, In Proceedings of The Fourth International Conference on Mobile Systems, Applications, and Services, Uppsala, Sweden, June 2006; pp. 220-232.

36. Linden, D.; Reddy, T. *Handbook of Batteries*, 3rd ed.; McGraw-Hill: New York, NY, USA, 2001.

37. Fuller, T. F.; Doyle, M.; Newman, J.; Simulation and Optimization of the Dual Lithium Ion Insertion Cell. J. Electrochemical Soc. **1994**, *141*, 1-10.

38. Chiasserini, C. F.; Rao, R. R. Pulsed battery discharge in communication devices. MobiCom '99, Seattle, WA, USA, August 1999; pp. 88-95.

39. Manwell, J.; McGowan, J.; Lead acid battery storage model for hybrid energy systems. *Solar Energy* **1993**, *50*, 399-405.

40. Martin, T. L. Balancing batteries, power, and performance: system issues in CPU speed-setting for mobile computing. Ph. D. Thesis, Carnegie Mellon University, Pittsburgh, Pennsylvania, USA, August 1999.

41. Rakhmatov, D.; Vrudhula, S. Energy management for battery-powered embedded systems. *ACM Trans. on Embedded Computing Systems* **2003**, *2*, 277-324.

42. Shih, E.; Bahl, P.; Sinclair, M. J. Wake on wireless: an event driven energy saving strategy for battery operated devices. In Proceedings of The 8th Annual International Conference on Mobile Computing and Networking (MobiCom '02), Atlanta, GA, USA, September 2002; pp. 160-171.

43. Agarwal, Y.; Schurgers, C.; Gupta, R. Dynamic power management using on demand paging for networked embedded systems. In Proceedings of The 2005 Conference on Asia and South Pacific Design Automation (ASP-DAC '05), Yokohama, Japan, January 2005; pp. 755-759.

44. Agarwal, Y.; Chandra, R.; Wolman, A.; Bahl, P.; Gupta, R. Wireless wakeups revisited: energy management for voip over wi-fi smartphones. In Proceedings of The 5th International Conference on Mobile Systems, Applications, and Services, San Juan, Puerto Rico, June 2007; pp. 179-191.

45. Brakmo, L. S.; Wallach, D. A.; Viredazand, M. A. μSleep: A Technique for Reducing Energy Consumption in Handheld Devices. In Proceedings of the 2nd International Conference on Mobile Systems, Applications, and Services, Zurich, Switzerland, June 2004; pp. 12-22.

46. Pathak, A.; Hu, Y. C.; Zhang, M.; Bahl, P.; Wang, Y.-M. Enabling automatic offloading of resource-intensive smartphone applications. In ECE Technical Reports, Purdue University, TR-ECE-11-13, May 2011.

47. Xian, C.; Lu, Y.-H.; Li, Z. Adaptive computation offloading for energy conservation on battery-powered systems. In Proceedings of International Conference on Parallel and Distributed Systems, Hsinchu, Taiwan, December 2007; pp. 1-8.

48. Kumar, K.; Lu, Y.-H. Cloud computing for mobile users: can offloading computation save energy. *Computer* **2010**, *43*, 51-56.

49. Abdesslem, F. B.; Phillips, A.; Henderson, T. Less is more: energy-efficient mobile sensing with senseless. In Proceedings of the 1st ACM Workshop on Networking, Systems, and Application for Mobile Handhelds, Barcelona, Spain, August 2009; pp. 61-62.

50. Kang, S.; Lee, J. Jang, H.; Lee, H.; Lee, Y.; Park, S.; Park, T.; Song, J. SeeMon: scalable and energy-efficient context monitoring framework for sensor-rich mobile environments. In Proceedings of the International Conference on Mobile Systems, Applications, and Services, Breckenridge, CO, USA, June 2008; pp. 267-280.

51. Priyantha, B.; Lymberopoulos, D.; Liu, J. Little rock: enabling energy-efficient continuous sensing on mobile phones. *IEEE Pervasive Computing* **2011**, *10*, 12-15.

52. Zhong, L.; Jha, N. K. Graphical user interface energy characterization for handheld computers. In Proceedings of the 2003 International Conference on compilers, Architecture and Synthesis for Embedded Systems, San Jose, CA, USA, October-November 2003; pp. 232-242.

53. Tiwari, V.; Malik, S.; Wolfe, A. Compilation techniques for low energy: an overview. IEEE Symposium on Low Power Electronics, San Diego, CA, USA, October 1994; pp. 38-39.

54. Aho, A. V.; Sethi, R.; Ullman, J. D. *Compilers, Principles, Techniques and Tools*; Addison Wesley: Boston, MA, USA, 1988.

55. Myers, B. A.; Rosson, M. B. Survey on user interface programming. Proc. of ACM Conference on Human Factors in Computing Systems, Monterey, CA, USA, May 1992; pp. 195-202.

56. Yan, X.; Sekercioglu, A.; Narayanan, S. A survey of vertical handover decision algorithms in Fourth Generation heterogeneous wireless networks. *Computer Networks* **2010**, *54*, 1848-1863.

57. Chamodrakas, I.; Martakos, D. A utility-based fuzzy TOPSIS method for energy efficient network selection in heterogeneous wireless networks. *Applied Soft Computing* **2011**, *11*, 3734-3743.

58. Chamodrakas, I.; Leftheriotis, I.; Martakos, D. In-depth analysis and simulation study of an innovative fuzzy approach for ranking alternatives in multiple attribute decision making problems based on TOPSIS. *Applied Soft Computing* **2011**, *11*, 900-907.

59. Ding, F.; Xia, F.; Zhang, W.; Zhao, X.; and Ma, C. Monitoring energy consumption of smartphones. In Proceedings of 1st International Workshop on Sensing, Networking, and Computing with Smartphones (PhoneCom), Dalian, China, October 2011; pp. 610-613.